\documentclass[journal]{IEEEtran}
\ifCLASSINFOpdf
\else
\fi
\usepackage{amsmath}
\usepackage{amssymb}
\usepackage{graphicx}
\usepackage{booktabs}
\usepackage[dvipsnames]{xcolor}
\usepackage[symbol]{footmisc}

\usepackage{pifont}
\newcommand{\cmark}{\ding{51}}%
\newcommand{\xmark}{\ding{55}}%

\begin{document}
%

\title{Learning Image Aesthetic Assessment from Object-level Visual Components}
%
%
%

\author{Jingwen Hou,~\IEEEmembership{Student Member,~IEEE,}
        Sheng Yang,
        ~Weisi Lin$^*$,~\IEEEmembership{Fellow,~IEEE,}
        Baoquan Zhao, and \\
        Yuming Fang,~\IEEEmembership{Senior Member,~IEEE,}
\thanks{J. Hou, S. Yang, W. Lin, B. Zhao are with the School of Computer Science and
Engineering, Nanyang Technological University, Singapore 639798. (e-mail:
jingwen003@e.ntu.edu.sg; syang014@e.ntu.edu.sg; wslin@ntu.edu.sg; bqzhao@ntu.edu.sg)}
\thanks{Y. Fang is with the School of Information Management, Jiangxi University
of Finance and Economics, Nanchang 330032, China (e-mail:
fa0001ng@e.ntu.edu.sg.}
\thanks{$^*$ Corresponding author.}
}

\maketitle

\begin{abstract}
As it is said by Van Gogh, great things are done by a series of small things brought together. Aesthetic experience arises from the aggregation of underlying visual components. However, most existing deep image aesthetic assessment (IAA) methods over-simplify the IAA process by failing to model image aesthetics with clearly-defined visual components as building blocks. As a result, the connection between resulting aesthetic predictions and underlying visual components is mostly invisible and hard to be explicitly controlled, which limits the model in both performance and interpretability. This work aims to model image aesthetics from the level of visual components. Specifically, object-level regions detected by a generic object detector are defined as visual components, namely object-level visual components (OVCs). Then generic features representing OVCs are aggregated for the aesthetic prediction based upon proposed object-level and graph attention mechanisms, which dynamically determines the importance of individual OVCs and relevance between OVC pairs, respectively. Experimental results confirm the superiority of our framework over previous relevant methods in terms of SRCC and PLCC on the aesthetic rating distribution prediction. Besides, quantitative analysis is done towards model interpretation by observing how OVCs contribute to aesthetic predictions, whose results are found to be supported by psychology on aesthetics and photography rules. To the best of our knowledge, this is the first attempt at the interpretation of a deep IAA model.

\end{abstract}



%
\IEEEpeerreviewmaketitle

\section{Introduction}

\IEEEPARstart{T}HE goal of image aesthetic assessment (IAA) is to evaluate the pleasantness of a photograph's aesthetic experience automatically. It can be used for a variety of tasks, such as image recommendation \cite{10.1145/2647868.2655053}, image editing \cite{wang2018deep}, image retrieval \cite{obrador2009role}, and photo management \cite{li2010towards}. 
In previous related works, three types of IAA tasks have been investigated: 1) binary classification, 2) score regression, and 3) rating distribution prediction. The binary classification task \cite{murray2012ava} splits images into high- and low-aesthetic categories, while the score regression task \cite{hosu2019effective} predicts average scores directly. A more difficult task, aesthetic rating distribution prediction (ARDP) \cite{talebi2018nima}, has recently drawn increased interest from the research community, not only because it better aligns with the uncertain nature of IAA, but also because the results of ARDP can be readily transformed to aesthetic scores or binary aesthetic labels. As the beauty is in the eye of the beholder, the ARDP allows for a genuine reflection of the nature of the IAA, i.e., the potential for greater diversity of agreement among the population on aesthetics.

Previous approaches resolve the problem of IAA by building aesthetic representation. Accordingly, these approaches can be roughly divided into three categories. The first category extracts features that directly correlate with image aesthetics according to photography rules, a.k.a hand-crafted feature-based approaches \cite{luo2011content, bhattacharya2010framework}. However, manually designing aesthetic representation is arduous. Benefited from deep learning, the second category learns aesthetic representation directly from images in an end-to-end manner \cite{talebi2018nima, xu2020context, chen2020adaptive}. And the third category learns aesthetic representation from generic descriptors of images in a data-driven manner, assuming that generic descriptors provide enough basic patterns for producing aesthetic representation \cite{marchesotti2011assessing, hosu2019effective}. Note that generic descriptors can be both low-level descriptors such as Bag-of-Visual-Words or Fisher Vector \cite{marchesotti2011assessing} and deep generic features such as MLSP features \cite{hosu2019effective}, though using deep generic features achieves better results.

However, building aesthetic representation is not a trivial task, since aesthetic assessment itself is a complex psychological process that has not yet been fully understood. Although promising results have been achieved by previous deep learning-based approaches, we argue that these approaches have over-simplified the problem of IAA by learning aesthetic representation directly from holistic images, ignoring the fact that aesthetic experience arises from the aggregation of visual components. In photography, the most important element of a photograph is the ``subject'', which is the most figurative object in the image. One standard for a good photograph is that the image should achieve \textbf{attention-subject consistency} \cite{tang2013content, sun2009photo, ke2006design, luo2008photo, freeman2007complete, freeman2007photographer}, meaning that the most salient object of a good photograph should be its subject.  Apart from the photography theory, psychological studies also suggest that a good organization of visual components should be \textbf{visually-right} \cite{locher2003empirical, nodine2008visual}, that is, there must be a right organization of pictorial objects and such an organization should be prominent to humans. Both theories imply: 1) aesthetic experience to the image originates from underlying visual components; 2) objects are basic visual components of a photograph; 3) attention on visual components or organizations of visual components are reflections of the pleasantness of aesthetic experience.

Our method is designed to better align with the aforementioned photography and psychology theories. First, we define visual components as object-level regions detected by a generic object detector, namely object-level visual components (OVCs). Second, we aggregate characteristics of OVCs into aesthetic representation via two attention mechanisms, including object-level attention mechanism and graph attention mechanism. The object-level attention mechanism determines the aesthetic contribution of individual OVCs, which mimics the process of finding the subjects. The object-level attention mechanism also complies with psychology studies with which basic units of human attention are visual objects \cite{scholl2001objects}. While the graph attention mechanism infers relevance between OVC pairs, which imitates the procedure of finding visually-right organizations of pictorial objects.
Inspired by Hosu \textit{et al.}'s work \cite{hosu2019effective}, we believe that deep generic features can be used to represent characteristics of OVCs, and therefore we further represent each visual component by a deep generic feature extracted from the associated OVCs. Then we produce aesthetic representation from those deep generic features according to learned attentions. 

We name our framework as Image Aesthetic Assessment from Object-level Visual Components (IAA-OVC). We believe that the benefits of the framework are four-fold. First, the object-level attention mechanism allows regional contributions to overall aesthetics to be determined in a finer granularity and with higher flexibility comparing to pixel-level attention since clear object boundaries have been introduced above pixels. Second, the graph attention mechanism enables explicit and dynamic determination of cross-region dependencies for non-local feature aggregation. Third, feature aggregation based on attentions allows the process for producing aesthetic representation to be explicitly controlled. Lastly, the combination of the object-level attention mechanism and the graph attention mechanism further enables the process of building aesthetic representation to be observed for model interpretation.

This work is an extended version of our paper published on ACMMM 2020 \cite{hou2020object}. The major difference is that we embed the object-level attention mechanism proposed in the previous version along with the graph-based attention mechanism into the framework of IAA-OVC. Additionally, quantitative analysis has been conducted on learned object-level and graph attentions with associated object labels including classes and attributes for deep IAA model interpretation. 
In a nutshell, our contributions can be summarized as follows:

\begin{itemize}
    \item Inspired by photography and psychology theories, we propose a new deep IAA paradigm for learning aesthetic representation from OVCs.
    \item We realize the model with the object-level attention mechanism and the graph attention mechanism for the explicit aggregation of generic descriptors representing OVCs. To the best of our knowledge, this is the first model that builds aesthetic representation based on the object-level attention mechanism and the graph attention mechanism.
    \item As far as we know, the proposed model is the first attempt towards an interpretable deep IAA method. We conduct extensive quantitative analysis on interpreting what has been learned by the model by associating each OVC with its object labels. We find that: 1) the model is able to distinguish subject regions from non-subject regions for some categories of photographs; 2) object-level attentions and graph attentions inferred by the trained model are negatively or positively correlate to aesthetic predictions. The results are found to be supported by the psychology and photography theories.
\end{itemize}

The remainder of the paper is organized as follows. The most related works is briefly introduced in Section II. We present the proposed IAA-OVC framework in Section III. Experimental results are given in Section IV. We present further analysis on model interpretation in Section V and we draw the conclusion in Section VI.

\begin{figure*}
  \centering
  \includegraphics[width=0.99\textwidth]{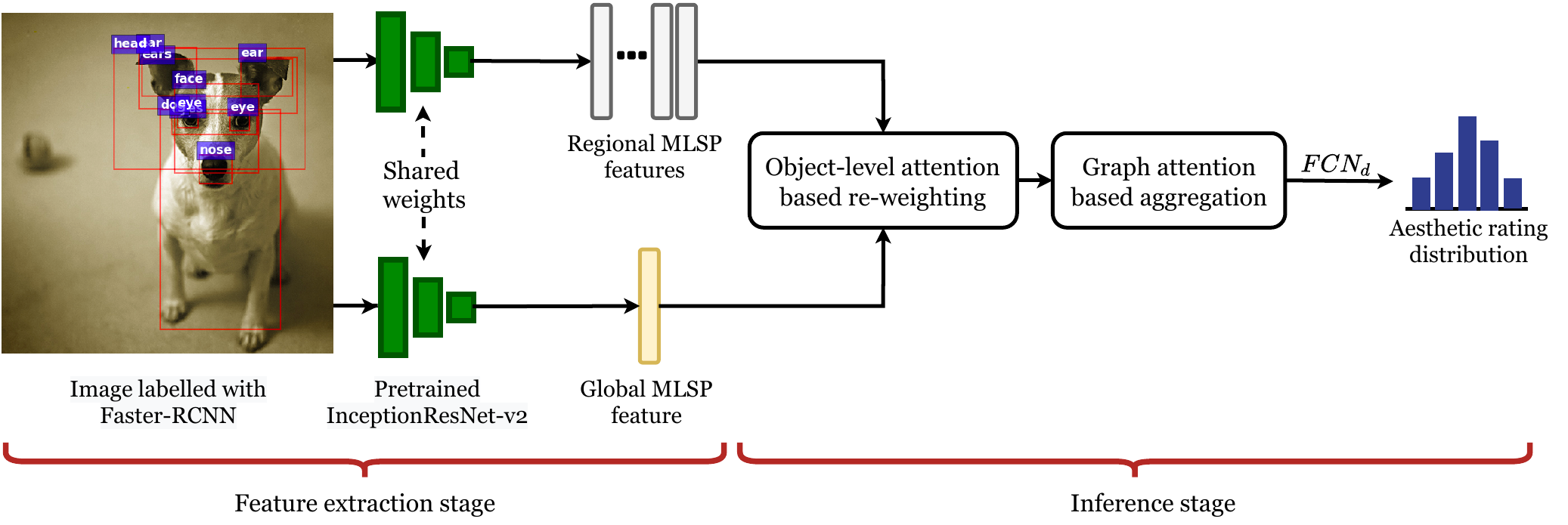}
  \caption{The diagram of the proposed framework. The model infers aesthetic rating distributions in two separate stages. In the first (feature extraction) stage, features are extracted with ImageNet pretrained InceptionResNet-v2 from a full-resolution image and object-level regions detected by a Faster-RCNN. In the second (inference) stage, with the help of the Object-level Attention-based Re-weighting module and Graph Attention-based Aggregation module, the model learns importance of individual object-level visual components and relevance between different object-level visual component pairs, and aggregates extracted features accordingly for inferring aesthetic rating distributions.}
  \label{fig:ran-overview}
\end{figure*}

\section{Related works}

IAA aims to automatically assess the pleasantness of the aesthetic experience of a given image. Early methods have designed features based on photography rules, in the hope that designed features would directly correlate with the pleasantness of aesthetic experience. For example, features for tone, colorfulness, luminance, composition, texture, sharpness, clarity, visual saliency \cite{datta2006studying, ke2006design, sun2009photo, bhattacharya2010framework, nishiyama2011aesthetic, luo2011content, zhang2014fusion} have been adopted. 
However, the effectiveness of these methods is limited, since high-level visual characteristics and interactions are hard to be represented in hand-crafted ways. More importantly, subject areas are hard to be defined manually due to the fact that subjects should be determined according to the context and any object or group of objects can be the potential subject of the image when multiple objects exist in the image. Compared to hand-crafted methods, the object-level attention mechanism proposed in our work is able to learn likelihoods of object-level regions to be the subject in a data-driven manner according to the context.


As more ground-truth data became available, later methods were designed to obtain aesthetic representation directly from generic features or images in a data-driven manner. Early data-driven approaches directly learned aesthetic representation from generic descriptors such as SIFT \cite{yeh2012relative}, Bag-of-Visual-Words and Fisher Vector \cite{marchesotti2011assessing}. However, the expressive power of these hand-crafted generic descriptors was quite limited which further leads to saturation of performance in IAA. Recently, Hosu \textit{et al.} \cite{hosu2019effective} proposed to use features extracted by ImageNet pretrained network as base features for building aesthetic representation in a data-driven manner, which presented a good performance on existing benchmarks. Our model is also designed to build aesthetic representation based on generic features. However, our method takes one step further. Instead of directly learning aesthetic representation from generic features extracted from the holistic image, generic features are adopted to describe characteristics of predefined OVCs, and aesthetic representation is generated by aggregating features representing OVCs.

Instead of learning from generic features, the other track of data-driven methods tried to obtain aesthetic representation directly from images in an end-to-end process. The pioneering work RAPID by Lu \textit{et al.} \cite{lu2014rapid} assumed that image aesthetics have to be learned from both a global and a local view (patches selected from the original-sized image), where the global view provides information about layout and composition while the local view provides fine-grained details. Along this line of thinking, a series of works continuously investigate more appropriate architecture for integrating composition or layout information and fine-grained details (e.g. image quality) in the process of aesthetic representation learning. Lu \textit{et al.} further proposed DMA-Net \cite{lu2015deep} that accepts 5 patches selected from holistic images as input, based on the assumption that any single patch is not able to represent the quality of fine-grained details for the whole image. However, since DMA-Net selected image patches randomly, the selected patches may not be diverse enough to represent the holistic image. Therefore, Ma \textit{et al.}  \cite{ma2017lamp} regard the patch selection process as an optimization problem: given an image and the number of patches to be selected, the selected patches should be as diverse as possible and with the least overlapping. Instead of using a heuristic approach as Ma \textit{et al.}'s work  \cite{ma2017lamp}, Zhang \textit{et al.} \cite{zhang2019gated} and Yang \textit{et al.} \cite{yang2019sgdnet} implement the patch selection as a learnable process by using spatial attention learned from global view to select a local patch from an original-sized image. 

Comparing to previous approaches \cite{lu2014rapid, lu2015deep, ma2017lamp}, our method enables a soft selection of patches from object-level candidates in an end-to-end and learnable manner. Instead of only choosing a fixed number of patches from the candidates, the object-level attention mechanism allows all candidates to be considered while with dynamically-determined contributions. Since candidates are at object-level instead of randomly-cropped at pixel-level, semantic integrity of selected regions are better maintained, and a finer granularity of regional contributions is possible to learn. Take portraits for example, the pixel-level attention mechanism \cite{zhang2019gated,yang2019sgdnet} only allows the face region to be highlighted, while the object-level attention mechanism can further distinguish contributions of different facial features. Apart from learning independent contributions of patches, the graph attention mechanism further allows contributions of patches to be modeled as groups and across regions.
Both attentions are inferred for explicitly aggregating features representing OVCs for aesthetic predictions, making it possible to observe the relation between aesthetic predictions and attended regions for model interpretation by cooperating attentions with associated object labels.

\section{Proposed Approaches}

\subsection{Problem Formulation}
Our approach focuses on the aesthetic rating distribution prediction (ARDP), since it is more closely aligned with the uncertain nature of human perception. The task of ARDP is to predict the aesthetic rating distribution (ARD) of a given image. Following the definition of the ARDP in previous works \cite{talebi2018nima, zeng2019unified}, we firstly formulate the problem as follows.
The raw ARD of the $i$-th image in the training set can be expressed as ${\mathcal{C}_i} = \{ c_i^j\} _{j = 1}^{K}$, where $c_i^j$ is the number of votes in the $j$-th score bucket of the $i$-th image and $K$ is the number of score buckets. In practice, we use AVA dataset and $K=10$ in this case. All raw ARD labels are normalized by being divided by the total number of votes in all buckets. Therefore, the normalized ARD of the $i$-th image is given by:

\begin{equation}
\label{eq:normalized_ard}
{\mathcal{P}_i} = \{ p_i^j\} _{j = 1}^{K} = \{ c_i^j/\sum\limits_{j = 1}^{K} {c_i^j} \} _{j = 1}^{K}, 
\end{equation}
then the training set with normalized ARD as labels can be denoted by $\{ (\mathbf{I}_i,{\mathcal{P}_i})\} _{i = 1}^N$, where $N$ is the number of images in the training set. If we denote the predicted ARD by ${\hat{\mathcal{P}}}_i$, then the optimization process for learning the parameter $\theta$ of the neural network model for ARDP can be expressed as:

\begin{equation}
\label{eq:ardp_opt}
\theta  = \mathop {\arg \min }\limits_\theta  \sum\limits_{i = 1}^N {\mathcal{L}({\mathcal{P}_i},{{\hat {\mathcal{P}}}_i})},
\end{equation}
where $\mathcal{L}(\cdot)$ is the loss function. For the loss function, we follow previous works \cite{talebi2018nima} and use the normalized Earth Mover Distance (EMD) loss. The normalized EMD loss is given by:
\begin{equation}
\label{eq:emd}
    {\mathcal{L}({\mathcal{P}_i},{{\hat {\mathcal{P}}}_i})} = \sqrt {{1 \over n}\sum\limits_{k = 1}^n {|{\rm CDF}{_{{\mathcal{P}_i}}}(k) - {\rm CDF}{_{{{\hat {\mathcal{P}}}_i}}}(k){|^2}} },
\end{equation}
where ${{\rm CDF}{_{{\mathcal{P}_i}}}}$ and ${{\rm CDF}{_{{\mathcal{\hat P}_i}}}}$ are cumulative density function for ground-truth ARD $\mathcal{P}_i$ and predicted ARD $\mathcal{\hat P}_i$, respectively, and $n$ is their length.

\subsection{Generic Feature Extraction}
\label{sec:feature_extraction}

We intend to construct aesthetic representation from generic features, assuming that aesthetic representation can be built upon fundamental patterns in generic features. To this end, the generic features selected should meet a number of criteria. First, considering IAA involves both low-level aspects such as image degradation and high-level aspects such as semantic information, the feature should contain responses to both low-level and high-level patterns. Second, the feature's high-level components should be able to cover a wide range of content so that the resulting IAA model is not limited to a single type of content.

According to aforementioned criteria, we employ multi-level spatially pooled (MLSP) \cite{hosu2019effective} features. 
MLSP features are extracted from InceptionResNet-v2 \cite{baldassarre2017deep} CNN pretrained on ImageNet.
An MLSP feature is constructed from outputs of different convolution blocks. The feature from each block is pooled into a fixed spatial size, and then the pooled features are sequentially concatenated into an MLSP feature. 
Two pooling strategies are proposed by the original work \cite{hosu2019effective}. 
The first one resized each feature map into $5\times5$, whereas the second one directly applies  global average pooling (GAP) to each feature map. The results of the first and the second strategies are referred to as Wide MLSP feature and Narrow MLSP feature, respectively.


We believe that the MLSP features satisfy our requirements for the following reasons. First, the model is pre-trained on ImageNet following a general classification task, and therefore a wide range of content are covered. Second, integrity of low-level details are better maintained since features are extracted from original-sized images. Third, since the concatenated feature integrates the outputs from each convolution block of the backbone model, it includes both low-level and high-level information. 

We use both regional and global MLSP features as basis for the construction aesthetic representation, as shown in Fig. \ref{fig:ran-overview}. Suppose the selected regions from image $\mathbf{I}$ is given as $\{ \mathbf{I}^i\} _{i = 1}^L$, where $L$ is the number of selected regions, the global and regional feature extraction is represented by:
\begin{equation}
    \label{eq:global_fe}
    {\mathbf{v}_{{\rm g}}} = {{M}_{\rm global}}({\mathbf{I}}),
\end{equation}
\begin{equation}
    \label{eq:regional_fe}
    {\mathcal{V}_{{\rm r}}} = \{ \mathbf{v}_{{\rm r}}^i\} _{i = 1}^L = {{M}_{\rm regional}}(\{ \mathbf{I}^i\} _{i = 1}^L),
\end{equation}
where ${\mathbf{v}_{{\rm g}}} \in {\mathbb{R}^{{{d}_{\rm g}}}}$ represents the global feature and $\mathbf{v}_{{\rm r}}^i \in {\mathbb{R}^{{{d}_{\rm r}}}}$ represents the regional feature extracted from the $i$-th selected region, and ${d}_{\rm g}$ and ${d}_{\rm r}$ are the length of the global and regional features, respectively. Both ${\mathbf{v}_{{\rm g}}}$ and ${\mathcal{V}_{{\rm r}}}$ are extracted with MLSP strategy but from different regions of the image $\mathbf{I}$:

\begin{figure}
  \centering
  \includegraphics[width=0.49\textwidth]{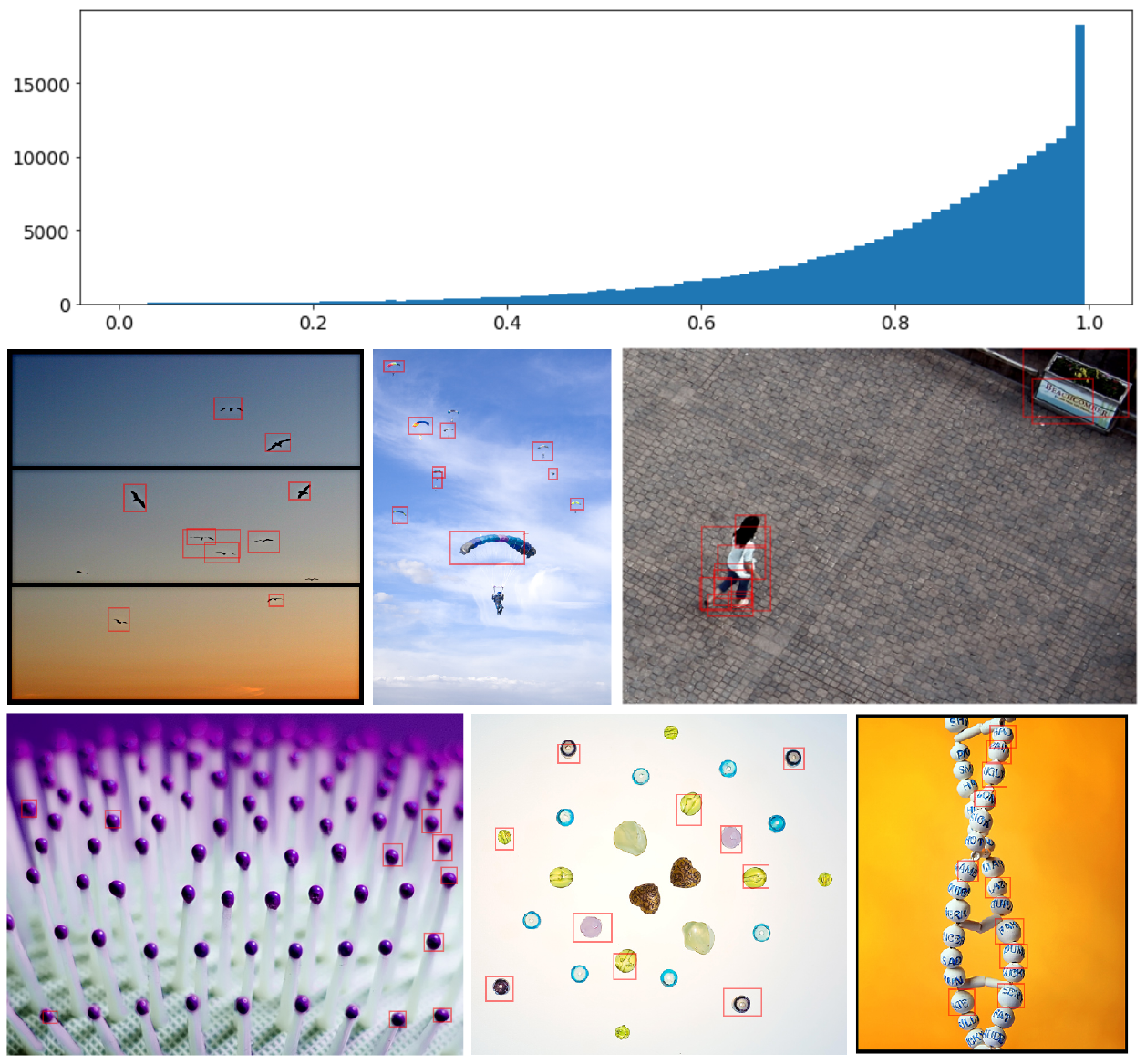}
  \caption{The demonstration of robustness of the selected object detector. Row 1: distribution of areas covered by bounding boxes of AVA dataset. Row 2: successful examples with proportion of covered regions $<$ 30\%. Row 3: failed examples with proportion of covered regions $<$ 30\%.}
  \label{fig:objdet}
\end{figure}

\textbf{Regional feature extraction}. 
Each regional feature corresponds to one of the OVCs defined by object-level RoIs. To be clear, object-level RoIs are object-level regions discovered by the chosen object detector. Specifically, Faster-RCNN \cite{ren2015faster} is employed as our object detector, as shown in Fig. \ref{fig:ran-overview}, since Faster-RCNN is the most widely used object detector \cite{anderson2018bottom, he2017mask, xu2017scene} in various scenarios. 
Choosing an appropriate dataset for training the Faster-RCNN is the next challenge. The main consideration is that it should cover a large diversity of objects, so that the granularity of detection can be as fine as possible. 
Though Faster-RCNN trained on Pascal VOC \cite{everingham2010pascal} or MS COCO \cite{lin2014microsoft} dataset are commonly-used versions, both of them still lack the granularity of detection that we desired. Therefore, we adopt the version trained on Visual Genome \cite{krishna2017visual}, which covers a much larger object categories than Pascal VOC \cite{everingham2010pascal} or MS COCO \cite{lin2014microsoft} (76,340 vs. 91 and 20). 
On image captioning and visual question answering \cite{anderson2018bottom, Yang_2019_CVPR, yang2020deconfounded}, the efficacy of this edition has been proven. The Faster-RCNN is then applied to all images in AVA dataset. We choose the top 10 region proposals for each image based on confidence because 10 regions are the maximum number of regions that can be handled by our hardware devices. While we believe 10 regions are able to cover most cases for the adopted dataset. Row 1 of Fig. \ref{fig:objdet} presents proportion of areas covered by bounding boxes of all images in AVA dataset, where the majority of images have more than 60\% of regions to be covered (82\% in average). This implies OVCs are able to be detected for most images. Even for abstract images, the chosen object detector can still define objects by their shapes or forms in most cases, e.g. line, circle, dot, rim, light, shadow, reflection, etc. We also present cases whose proportion of covered regions is less than 30\%. While less covered regions does not mean to be failure, since some cases have less subjects on a clear background as shown in Row 2 of Fig. \ref{fig:objdet}. We do observe cases can be regarded as failure. These cases are typically repetitive presence of a similar pattern as shown in Row 3 of Fig. \ref{fig:objdet}, while these cases are very rare and can be solved by increasing the number of regions to be considered. 

\textbf{Global feature extraction}. 
The global features extracted from original-sized images are adopted to provide global information such as image layout and compensate for information from regions that have not been covered by object-level RoIs. We consider both narrow and wide MLSP settings for global feature extraction, and their performances are discussed in Section \ref{sec:ablation_study_gaa}.

\begin{table}[]
\caption{Specification and notation of fully-connected networks used in the proposed model}
\label{table:fcn_spec}
\centering
\begin{tabular}{l|l|l}
\toprule
\textbf{Notation} & \textbf{Notes}                                                                                              & \textbf{Specification}   \\
\toprule
$\operatorname{FCN}_{\rm g}$   & \begin{tabular}[c]{@{}l@{}}Global MLSP feature\\ dimensional reduction\end{tabular}                & {[}FC(16928, 6144), ReLU{]}   \\
\midrule
$\operatorname{FCN}_{\rm r}$   & \begin{tabular}[c]{@{}l@{}}Regional feature\\ dimensional reduction\end{tabular}                   & {[}FC(16928, 256), ReLU{]}  \\
\midrule
$\operatorname{FCN}_{\rm o}$   & \begin{tabular}[c]{@{}l@{}}Object-level attention\\ predictor\end{tabular}                       & \begin{tabular}[c]{@{}l@{}}{[}FC(8704, 4096), BN, ReLU{]}\\ {[}FC(4096,10), Sigmoid{]}\end{tabular} \\
\midrule
$\operatorname{FCN}_{\rm d}$   & Distribution predictor                                                                             & {[}FC(2560, 10), Softmax{]}   \\
\midrule
$\operatorname{FCN}_{\rm g'}$  & \begin{tabular}[c]{@{}l@{}}Global feature\\ dimensional reduction \\ for node feature\end{tabular} & {[}FC(6144, 128), ReLU{]}   \\
\midrule
$\operatorname{FCN}_{\rm e}$   & \begin{tabular}[c]{@{}l@{}}Visual component\\ relevance predictor\end{tabular}                     & \begin{tabular}[c]{@{}l@{}}{[}FC(260, 260), ReLU{]} \\ {[}FC(260, 1){]}     \end{tabular} \\
\bottomrule
\end{tabular}
\end{table}

\subsection{Object-level Attention-based Re-weighting (OAR)}
\label{sec:network}

\begin{figure}
  \centering
  \includegraphics[width=0.48\textwidth]{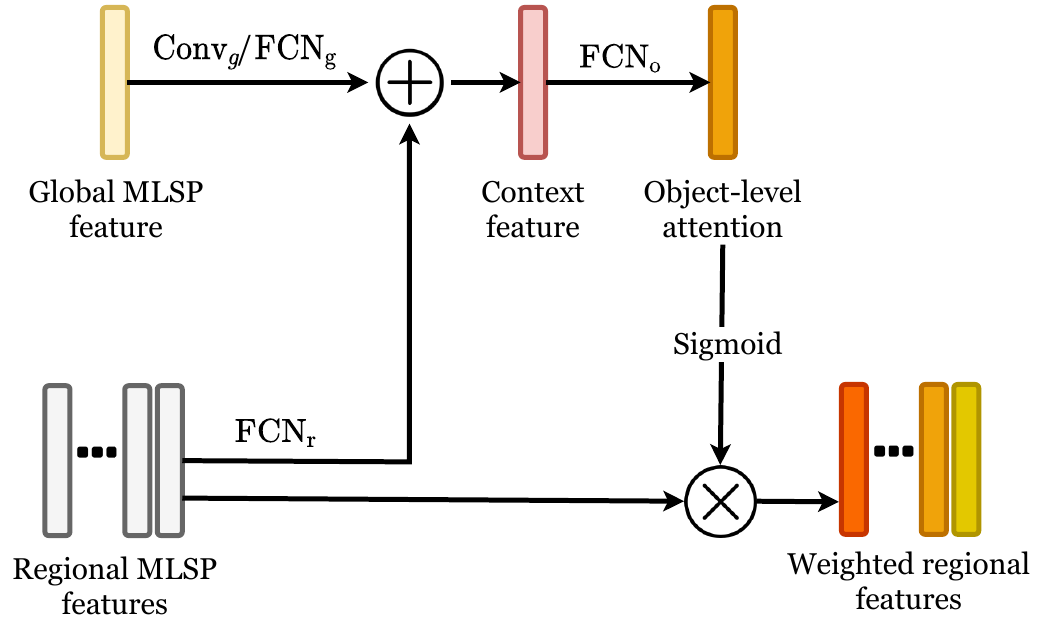}
  \caption{The architecture of the object-level attention-based re-weighting module.}
  \label{fig:obj_att}
\end{figure}

Object-level attentions are firstly learned to re-weight regional features. This process is implemented as the module named Object-level Attention-based Re-weighting (OAR), whose architecture is shown in Fig. \ref{fig:obj_att}. Taking a global feature and a set of regional features as input, the OAR module re-weights regional features for the next step. Given a global feature $\mathbf{v}_{\rm g}$ and a set of regional features $\{ \mathbf{v}_{{\rm r}}^i\} _{i = 1}^L$, the dimensions of features are first reduced by the global and regional feature dimensional reduction (FDR) modules. 
Inspired by \cite{hosu2019effective}, the module for the wide MLSP setting for obtaining the global feature (denoted as ${\rm Conv}_{\rm g}$ in Table \ref{table:fcn_spec}) is constructed with 3 different convolution blocks, where each block reduces the channel dimension of wide MLSP features from 16928 to 2048 while preserving its size of spatial dimension. Thus, by concatenating the outputs of the 3 convolution blocks, the $5\times5\times16928$ feature is shrunk to $5\times5\times6144$. Then its spatial dimension is further dwindled by global average pooling (GAP), which generates the final $1\times1\times6144$ dimension-reduced global feature. When employing the narrow MLSP setting, the FDR modules for global features or regional features are based on fully-connected networks (FCNs). The specification of FCN-based FDR modules is shown in Table \ref{table:fcn_spec} (denoted as $\operatorname{FCN}_{\rm r}$ and $\operatorname{FCN}_{\rm g}$ for regional and global FDR module, respectively). After dimensional reduction, the size of a global feature is reduced to 6144 and the size of each regional feature is reduced to 256. 


Because there are $L$ RoIs for each image, there are $L$ corresponding regional features. Suppose $\mathcal{V}' = \{ {\mathbf{v}^i}\} _{i = 1}^{L + 1}$ is the set of dimension-reduced global and regional features, where $\mathbf{v}^{L+1} \in \mathbb{R}^{6144}$ is the dimension-reduced global feature and the remaining are dimension-reduced regional features $\mathbf{v}^{i} \in \mathbb{R}^{256}$ and $i \in [1, L]$. Since we take $L=10$ in practice, all features are concatenated into one 8704-length feature vector and passed to the attention predictor. The target of the attention predictor is to predict the contribution of each regional feature, and we realize the attention predictor with an FCN that takes an 8704-length input and produces a 10-length attention vector $\mathbf{a}$. The specification of the FCN-based attention predictor $\rm {FCN}_{o}$  is presented in Table \ref{table:fcn_spec}. Because the attention predictor adopts sigmoid activation at its output layer, the scale of predicted weights is $(0, 1)$. The inference of the attention weights can be summarized as:
\begin{equation}
    \mathbf{a} = {\rm {FCN}_{o}}(\mathop  \oplus \limits_{i = 1}^{L + 1} \mathbf{v}^i),
\end{equation}
where $\oplus$ represents the concatenation operation and $ \mathbf{a} \in \mathbb{R}^L$ is the attention vector predicted by the attention predictor $f_{\rm attention}(\cdot)$. Finally, each of the regional features is weighted by applying corresponding attention weight:

\begin{equation}
{\widetilde{\mathcal{V}} = \{ \widetilde{\mathbf{v}}^i\} _{i = 1}^L = \{ { \mathbf{a}_i} \cdot \mathbf{v}^i\} _{i = 1}^L},
\end{equation}
where $\widetilde{\mathcal{V}}$ is the set of weighted regional features. Finally, the weighted regional features are further concatenated and passed to the FCN-based distribution predictor (denoted as $\rm {FCN}_d$ in Table \ref{table:fcn_spec}) to infer the final ARD.

\subsection{Graph Attention-based Aggregation (GAA)}
\label{sec:gaa}

\begin{figure}
  \centering
  \includegraphics[width=0.5\textwidth]{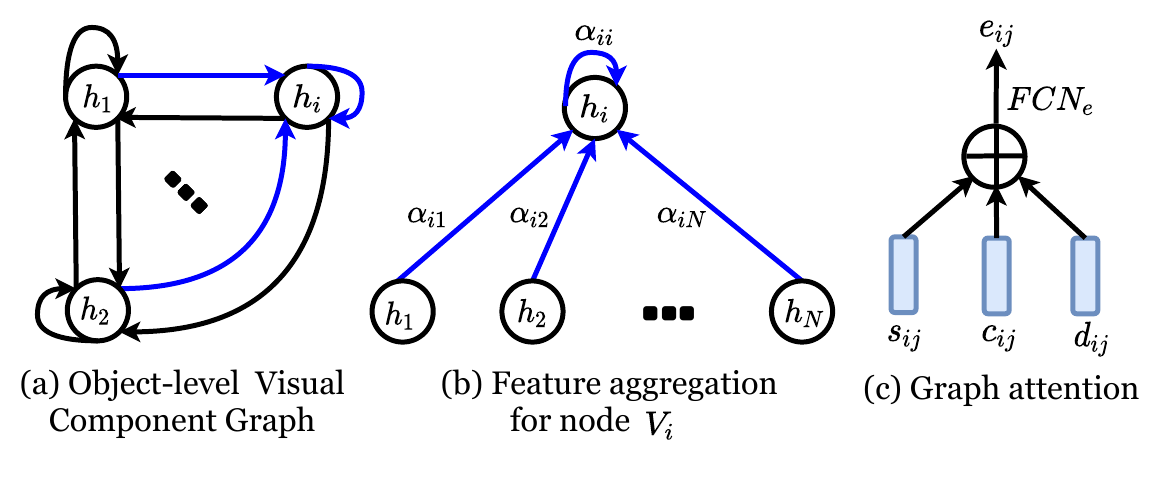}
  \caption{The demonstration for key concepts in our graph attention-based aggregation. Connections to a single node is shown in \textcolor{blue}{blue}. }
  \label{fig:graph_att}
\end{figure}

The graph attention-based aggregation (GAA) is inspired by \cite{velivckovic2018graph}. Each image $\mathbf{I}$ is parsed into an Object-level Visual Component Graph (OVCG) for ARDP. The OVCG is constructed to capture pairwise relevance between OVCs. The OVCG $\mathcal{G}=\{\mathcal{V}, \mathcal{E}\}$ is composed of a node set $\mathcal{V}$ and an edge set $\mathcal{E}$. Each node $V_i \in \mathcal{V}$ represents an OVC and is associated with a node feature $\mathbf{h}_i$:
\begin{equation}
    \mathbf{h}_i = [\widetilde{\mathbf{v}}^i \|  \operatorname{FCN}_{\rm g'} \circ \operatorname{Conv}_{\rm g}(\mathbf{v}_{\rm g})],
\end{equation}
where node feature $\mathbf{h}_i \in \mathbb{R}^{384}$ in practice, and $\operatorname{FCN}_{\rm g'}(\cdot)$ (shown in Table \ref{table:fcn_spec}) reduces length of global feature $\mathbf{v}_{\rm g}$ into $128$, and $\|$ denotes concatenation between two features. For each visual component, it is possible that any visual component in the image could interact with it. To abstract this connection in the OVCG, each node in the OVCG is connected to all nodes including itself with unidirectional connections, as shown in Fig. \ref{fig:graph_att}(a).

For each node in the OVCG, we aim to model the relevance between this node and all nodes in the same OVCG. We first consider the case for the subgraph $\mathcal{G}_i=\{\mathcal{V}, \mathcal{E}_i\}$, where $\mathcal{E}_i = \{E_{ij}\}_{j=1}^N$ denotes all edges directing to the node $V_i$, $\mathcal{V} = \{V_{k}\}_{k=1}^N$ denotes all nodes in the OVCG, and $N$ is the number of all nodes in the OVCG $\mathcal{G}$. A demonstration is shown in Fig. \ref{fig:graph_att}(b). We call node $V_i$ as the central node of subgraph $\mathcal{G}_i$. Then the relevance is determined between each node in $\mathcal{V}$ and the central node $V_i$ via the graph attention mechanism. We consider 3 major aspects for modeling such relevance, including visual characteristic relationship, semantic relationship, and spatial relationship:

\textbf{Visual characteristic relationship.} Visual characteristic relationship involves contrast between visual characteristics of OVCs. As to visual characteristics, we mainly refer to low-level characteristics such as sharpness, color, luminance, texture, shape, and high-level patterns such as the comparison between semantics. The main reason for considering visual characteristic relationship for determining relevance between OVC pairs is that visual characteristics themselves can express significant aesthetics. For example, the repetitive occurrence of an appealing pattern, or a strong contrast among colors could make a photograph attractive. Since the adopted deep features itself contain responses to diverse complex patterns, the visual character relationship can be described by similarities between a central node feature and all node features. For a pair of nodes, the visual character relationship is given by:
\begin{equation}
    s_{ij} = \| {\operatorname{tanh}(\mathbf{W}_{\rm att}\mathbf{h}_i) - \operatorname{tanh}(\mathbf{W}_{\rm att}\mathbf{h}_j)} \|_1,
\end{equation}
where $\mathbf{W}_{\rm att} \in \mathbb{R}^{128 \times 384}$ is the weights for dimensional reduction.

\textbf{Semantic relationship.} Semantic relationship refers to the co-occurrence of high-level semantics. When certain semantic information is recognized from a photograph, people tend to expect other information to occur along with it. For example, when we see birds, we may expect a blue sky or a lush forest. Thus, the semantic relationship is another important aspect for the representation learning for aesthetics evaluation. To capture semantic relationship from co-occurrence for a pair of nodes, we adopt the non-linear mapping of concatenated node features:
\begin{equation}
    \mathbf{c}_{ij} = [\operatorname{tanh}(\mathbf{W}_{\rm att}\mathbf{h}_i) \| \operatorname{tanh}(\mathbf{W}_{\rm att}\mathbf{h}_j)],
\end{equation}
where $\mathbf{c}_{ij} \in \mathbb{R}^{256}$.

\textbf{Spatial relationship.} Spatial relationship depicts the layout of OVCs in a photograph. The layout of OVCs is also important in aesthetics, since different layout can direct the scanpath of observers and even express different feelings. For example, adopting the rule-of-third can direct the attention of observers to the object on the stress point, and a diagonal composition can make the object on the diagonal feeling stretched. Although there are many composition rules have been well studied in theories of photography, there are cases of high aesthetics that do not align with any of the existing rules, which implies there could be more patterns of such spatial relationship to be investigated. This leads us to consider spatial relationship. Given the coordinates of top-left and bottom-right corners of two bounding boxes of two nodes $p_{i}^{\rm tl}=(x_{i}^{\rm tl}, y_{i}^{\rm tl})$, $p_{i}^{\rm br}=(x_{i}^{\rm br}, y_{i}^{\rm br})$, $p_{j}^{\rm tl}=(x_{j}^{\rm tl}, y_{j}^{\rm tl})$ and $p_{j}^{\rm br}=(x_{j}^{br}, y_{j}^{\rm br})$, the spatial relationship is obtained by:

\begin{itemize}
    \item Euclidean distance between centers of bounding boxes of OVCs $d_{ij}^{\rm E}$:
    \begin{equation}
        \begin{gathered}
          p_{i}^{\rm center} = (\frac{{x_{i}^{\rm tl} + x_{i}^{\rm br}}}{2},\frac{{y_{i}^{\rm tl} + y_{i}^{\rm br}}}{2}) \hfill, \\
          p_{j}^{\rm center} = (\frac{{x_{j}^{\rm tl} + x_{j}^{\rm br}}}{2},\frac{{y_{j}^{\rm tl} + y_{j}^{\rm br}}}{2}) \hfill, \\
          d_{ij}^{\rm E} = \| p_{i}^{\rm center} - p_{j}^{\rm center}\|_2.
        \end{gathered} 
    \end{equation}
    \item Bi-directional Hausdorff distance between the regions of bounding boxes of OVCs $d_{ij}^{\rm H}$. The directed Hausdorff distance $\Tilde{d}_{ij}^{\rm H}$ between point sets $\mathcal{S}_i$ and $\mathcal{S}_j$ of the bounding box regions of the $i$-th and the $j$-th object is the maximum of distances between each point $p_i \in \mathcal{S}_i$ to its nearest neighbors $p_j \in \mathcal{S}_j$. Since $\Tilde{d}_{ij}^{\rm H} \ne \Tilde{d}_{ji}^{\rm H}$, the bi-directional Hausdorff distance between $\mathcal{S}_i$ and $\mathcal{S}_j$ is given by the maximum between $\Tilde{d}_{ij}^{\rm H}$ and $\Tilde{d}_{ji}^{\rm H}$ \cite{taha2015efficient}:
    \begin{equation}
        \begin{gathered}
            \Tilde{d}_{ij}^{\rm H} = \operatorname{max}_{p_i \in \mathcal{S}_i}\operatorname{min}_{p_j \in \mathcal{S}_j}\{\|\ p_i, p_j\|_2\}, \\
            \Tilde{d}_{ji}^{\rm H} = \operatorname{max}_{p_j \in \mathcal{S}_j}\operatorname{min}_{p_i \in \mathcal{S}_i}\{\|\ p_i, p_j\|_2\},
        \end{gathered}
    \end{equation}
    \begin{equation}
        d_{ij}^{\rm H} = \operatorname{max}\{\Tilde{d}_{ij}^{\rm H}, \Tilde{d}_{ji}^{\rm H}\}.
    \end{equation}
    \item IoU between areas of bounding boxes of OVCs  $d_{ij}^I$.
    \begin{equation}
        d_{ij}^{\rm I} = \frac{\mathcal{B}_i \cap \mathcal{B}_j}{\mathcal{B}_i \cup \mathcal{B}_j},
    \end{equation}
    where $\mathcal{B}_i$ and $\mathcal{B}_j$ are areas bounded by bounding boxes of the $i$-th and the $j$-th object. 
\end{itemize}

\begin{table*}[]
    \centering
    \caption{Peer comparison on AVA benchmark. Top 2 results on each metric are shown in \textbf{bold face.}}
    \begin{tabular}{l|cc|cc|c}
    \toprule
    Method                                             & SRCC (mean) & PLCC (mean) & SRCC (std. dev) & PLCC (std. dev) & Accuracy\\
    \midrule
    Talebi et al. (TIP 2018) \cite{talebi2018nima}                           & 0.636       & 0.612       & 0.233           & 0.218      & 81.51\%     \\
    Zhang et al. (TMM 2019)   \cite{zhang2019gated}                          & 0.690       & 0.704      & -               & -        & 81.81\%       \\
    Hosu et al.$^*$ (CVPR 2019) \cite{hosu2019effective}                          & 0.740       & 0.742       & 0.333           & 0.344    & 80.97\%       \\
    Zeng et al. (TIP 2020) \cite{zeng2019unified}                            & 0.719       & 0.720       & 0.241           & 0.247      & 80.81\%      \\
    Xu et al. (ACMMM 2020) \cite{xu2020context}      & 0.724      & 0.725       & -           & -   & 80.90\%         \\
    Chen et al. (CVPR 2020) \cite{chen2020adaptive}      & 0.649       & 0.671      & -           & -   & \textbf{83.24\%}         \\
    \midrule
    Ours (ACMMM 2020 version) \cite{hou2020object}     & \textbf{0.751}       & \textbf{0.753}       & \textbf{0.353}           & \textbf{0.363}   & 81.67\%         \\
    Ours (Extended version)       & \textbf{0.756}       & \textbf{0.757}       & \textbf{0.362}           & \textbf{0.374}   & \textbf{81.93\%}         \\
    \bottomrule
    \end{tabular}
    \label{tab:peer}
    \vspace{1mm}

   $^*$ The method has been retrained on the aesthetic rating distribution task.  
\end{table*}

Then by combining all above-mention measurements, we have the features for describing spatial relationship:
\begin{equation}
    \mathbf{d}_{ij} = [d_{ij}^{\rm E} \| d_{ij}^{\rm H} \| d_{ij}^{\rm I}],
\end{equation}
where $\mathbf{d}_{ij} \in \mathbb{R}^3$.

Finally, features of visual characteristic relationship, semantic relationship, and spatial relationship are combined for modeling graph attentions over the subgraph $\mathcal{G}_i$:
\begin{equation}
    e_{ij} = \operatorname{FCN}_{e}([s_{ij} \| \mathbf{c}_{ij} \| \mathbf{d}_{ij}]),
\end{equation}
\begin{equation}
    \alpha_{ij} = \operatorname{softmax}_j(e_{ij}),
\end{equation}
where the specification of $\operatorname{FCN}_{\rm e}(\cdot)$ is shown in Table \ref{table:fcn_spec}. Then we aggregate features of each subgraph with attention, which produces the aesthetic representation for node $V_i$:
\begin{equation}
    \mathbf{h}_i^{'} = \operatorname{eLU} (\sum\limits_{j = 1}^N {\alpha_{ij}[\mathbf{W}_{\rm agg}{\mathbf{h}_j}]} ),
\end{equation}
where $\mathbf{W}_{\rm agg} \in \mathbb{R}^{256 \times 384}$ are learnable parameters for reducing dimensions of node features.

\section{Experiment and analysis}

\subsection{Experimental Setup}

\begin{itemize}

\item \textbf{Datasets.} Following previous works \cite{mai2016composition, lu2015deep, hosu2019effective, zhang2019gated, li2020personality, zeng2019unified}, our experiments are performed on the official split of AVA dataset. It consists of approximately $250,000$ images, and officially divided into a training set with approximately $230,000$ and a testing set with approximately $20,000$ images. The dataset provides ground truth ratings for each image in form of raw rating distribution and average scores on a scale of $1\sim10$. Because some images are not available, there are 235,574 images for training and $19,928$ images for testing in practice. AVA dataset has also been officially divided into 9 subsets according to image contents, including ``floral'', ``animal'', ``portrait'', ``architecture'', ``landscape'', ``cityscape'', ``fooddrink'', ``still life'' and ``generic''. The generic subset contains approximately $20,000$ training images and approximately $20,000$ testing images of various contents, while each of the other 8 subsets has approximately $2,500$ training images and approximately $2,500$ testing images of specific contents.

\item \textbf{Evaluation.} The evaluation on ARDP task follows \cite{talebi2018nima}. Ground truth ARD and predicted ARD results are converted to average scores and standard deviations. Given a normalized ARD $\{ {p^j}\} _{j = 1}^{10}$, the average score is computed as $\mu  = \sum\limits_{j = 1}^{10} {j \cdot {p^j}} $ and the standard deviation is computed as $\sigma  = \sqrt {\sum\limits_{j = 1}^{10} {{{(j - \mu )}^2} \cdot {p^j}} } $. 
We adopt two commonly-used metrics, Spearman rank order coefficient (SRCC) and Pearson linear correlation coefficient (PLCC), for evaluating the goodness of fitting of average scores and standard deviations. We also convert ARD to binary labels with 5 as the cut-off threshold as previous works \cite{murray2012ava} and then accuracy is computed. 

\item \textbf{Training settings}. Our models in all settings in our experiments are trained on the official training set for 10 epochs with Adam optimizer and batch size 128 and then evaluated on the testing set. The learning rate is set to 3e-5 for the first 2 epochs and divided by 10 every 3 epochs.

\end{itemize}


\begin{table*}[]
    \centering
    \caption{Ablation study on the IAA-OVC framework on AVA benchmark. Best results are shown in \textbf{bold face}.}
    \begin{tabular}{l|cc|cc|c}
    \toprule
    Setting     & SRCC (mean) & PLCC (mean) & SRCC (std. dev) & PLCC (std. dev) & Accuracy \\
    \midrule
    Baseline                        & 0.652       & 0.654       & 0.225           & 0.233           & 78.10\%  \\
    OAR   & 0.751       & 0.753       & 0.353           & 0.363           & 81.67\%  \\
    OAR+GAA   & \textbf{0.756}       & \textbf{0.757}       & \textbf{0.362}           & \textbf{0.374}   & \textbf{81.93}\%         \\
    \bottomrule
    \end{tabular}
    \label{tab:ablation_arrl}
\end{table*}

\begin{table*}[]
    \centering
    \caption{Ablation study on the IAA-OVC framework on sub-categories of AVA dataset.  Best results are shown in \textbf{bold face}.}
    \begin{tabular}{l|l|cc|cc|c}
    \toprule
    Category     & Setting  & SRCC (mean) & PLCC (mean) & SRCC (std. dev) & PLCC (std. dev) & Accuracy     \\
    \midrule
                 & Baseline & 0.322     & 0.312     & 0.022     & 0.028     & 70.17\% \\
    Floral       & OAR       & 0.647     & 0.637     & 0.225     & 0.233     & 76.82\% \\
                 & OAR+GAA    & \textbf{0.669}     & \textbf{0.662}     & \textbf{0.253}     & \textbf{0.243}     & \textbf{77.71\%} \\
    \midrule
                 & Baseline & 0.343     & 0.345     & 0.091     & 0.092     & 70.85\% \\
    Animal       & OAR       & 0.626     & 0.629     & 0.208     & 0.198     & 76.61\% \\
                 & OAR+GAA    & \textbf{0.646}     & \textbf{0.657}     & \textbf{0.223}     & \textbf{0.230}     & \textbf{77.54\%} \\
    \midrule
                 & Baseline & 0.281     & 0.280     & 0.015     & 0.022     & 75.68\% \\
    Portrait     & OAR       & 0.593     & 0.594     & \textbf{0.204}     & \textbf{0.198}     & 79.10\% \\
                 & OAA+GAA    & \textbf{0.607}     & \textbf{0.614}     & 0.195     & 0.192     & \textbf{80.79\%} \\
    \midrule
                 & Baseline & 0.275     & 0.273     & 0.072     & 0.077     & 74.15\% \\
    Architecture & OAR       & 0.632     & 0.625     & 0.133     & 0.139     & 78.96\% \\
                 & OAR+GAA    & \textbf{0.659}     & \textbf{0.660}     & \textbf{0.150}     & \textbf{0.152}     & \textbf{79.36\%} \\
    \midrule
                 & Baseline & 0.331     & 0.325     & 0.048     & 0.051     & 77.31\% \\
    Landscape    & OAR       & 0.671     & 0.673     & 0.190     & 0.200     & 81.97\% \\
                 & OAR+GAA    & \textbf{0.688}     & \textbf{0.697}     & \textbf{0.209}     & \textbf{0.225}     & \textbf{82.93\%} \\
    \midrule
                 & Baseline & 0.388     & 0.394     & 0.016     & 0.021     & 71.21\% \\
    Cityscape    & OAR       & 0.664     & 0.655     & 0.146     & 0.147     & 77.67\% \\
                 & OAR+GAA    & \textbf{0.689}     & \textbf{0.685}     & \textbf{0.164}     & \textbf{0.169}     & \textbf{79.35\%} \\
    \midrule
                 & Baseline & 0.377     & 0.380     & -0.040    & -0.038    & 67.58\% \\
    Fooddrink    & OAR       & 0.643     & 0.644     & 0.128     & 0.125     & 76.04\% \\
                 & OAR+GAA    & \textbf{0.663}     & \textbf{0.673}     & \textbf{0.147}     & \textbf{0.133}     & \textbf{76.61\%} \\
    \midrule
                 & Baseline & 0.239     & 0.239     & -0.005    & 0.004     & 62.75\% \\
    Still life   & OAR       & 0.560     & 0.566     & 0.110     & 0.110     & 72.22\% \\
                 & OAA+GAA    & \textbf{0.581}     & \textbf{0.594}     & \textbf{0.134}     & \textbf{0.137}     & \textbf{73.18\%} \\
    \bottomrule
    \end{tabular}
    \label{tab:ablation_arrl_subsets}
\end{table*}

\subsection{Peer Comparison}

We compare our model with previous relevant works. We compare our results in the previous conference version \cite{hou2020object} and the updated version to 6 state-of-the-art relevant works \cite{talebi2018nima, zhang2019gated, zeng2019unified, xu2020context, chen2020adaptive, hosu2019effective} published on top conferences and journals. Note that for the work by Hosu \textit{et al.} \cite{hosu2019effective}, since the original work is trained subject to score regression and we have directly adopted their features in our own model, we have modified the model by replacing its output layer with a 10-way softmax layer and re-trained it with normalized EMD loss (Eq. (\ref{eq:emd})) on the ARDP task with AVA trainset for a fair comparison. For other works, we take the results from the original paper. Since other contenders \cite{talebi2018nima, zhang2019gated, zeng2019unified, xu2020context, chen2020adaptive} are trained subject to the ARDP task, we directly take their reported results. The evaluation results on AVA testset of the altered version are reported. All results are presented in Table \ref{tab:peer} and the top 2 results on each measurement are shown in \textbf{bold face}.

As shown, we can observe our model outperforms all selected methods in terms of SRCC and PLCC of average scores and standard deviations. This indicates our model is superior to other selected models in ARDP. We also notice that our accuracy is not the best among all presented methods. We argue that accuracy cannot well reflect the goodness of an IAA model as discussed in \cite{hosu2019effective}, because accuracy is very sensitive to predictions around the cut-off threshold, while insensitive to predictions far away from the threshold. For example, the scores 5.01 and 9 can both fall into the high aesthetic class when we set 5 as the cut-off threshold although these are two distinct scores, while the scores 5.01 and 4.99 can fall into two different classes although these two scores are very close. Additionally, accuracy also cannot reflect the uncertainty of agreement on visual aesthetics, contrary to ARD which can reflect the concentration of ratings.

\subsection{Ablation Study on the IAA-OVC Framework}


\label{sec:ablation_study_arrl}
We conduct an ablation study to investigate the effectiveness of the Object-level Attention-based Re-weighting (OAR) module and the Graph Attention-based Aggregation (GAA) module under the IAA-OVC framework. For the baseline setting, we directly concatenate all regional features with equal weights for learning on aesthetic assessment. Then we upgrade the baseline model by introducing the OAR module to dynamically assign weights to regional features, and aggregate features with graph attentions via the GAA module. All settings are evaluated on AVA benchmark and content-specific subsets of AVA dataset. The results on AVA benchmark are shown in Table \ref{tab:ablation_arrl} and the results on AVA subsets are shown in Table \ref{tab:ablation_arrl_subsets}. We can observe all measurements are improved as the model setting is upgraded. This supports our assumption that the OAR and the GAA module can help to better integrate features of individual OVCs than learning with traditional fully-connected networks. This also implies generic network designs, e.g. fully-connected architecture, may not suit the best for learning aesthetic representation and the proposed attention mechanisms help for better feature aggregation. Similar conclusions can also be derived from the results on each contents-specific subset (Table \ref{tab:ablation_arrl_subsets}), which confirms the generalizability of our framework on different scenarios of IAA. While we also observe that the improvement brought by GAA is much less than OAR, we speculate that it is because OAR along with the rest of the architecture already allows pairwise interactions to be modeled implicitly and therefore less extra information is learned by GAA for further improvements. Though the GAA module has not brought significant quantitative performance, it enables pairwise relevance to be observed for model interpretation, which will be further discussed in Sec. \ref{sec:gatt_vs_score}.

\begin{table*}[]
\centering
\caption{Ablation study on OAR module on generic subset of AVA dataset.  Best results are shown in \textbf{bold face}.}
\begin{tabular}{l|cc|cc|c}
\toprule
Setting                & SRCC (mean) & PLCC (mean) & SRCC (Std. dev) & PLCC (Std. dev) & Accuracy     \\
\midrule
Baseline               & 0.557     & 0.561     & 0.082    & 0.085    & 75.25\% \\
Regional               & 0.597     & 0.599     & 0.156    & 0.160    & 76.54\% \\
Regional+Narrow global & 0.683     & 0.686     & 0.240    & 0.247    & 79.43\% \\
Regional+Wide global   & \textbf{0.701}     & \textbf{0.704}     & \textbf{0.265}    & \textbf{0.272}    & \textbf{80.00\%} \\
\bottomrule
\end{tabular}
\label{tab:abl_oar}
\end{table*}

\begin{table*}[]
\centering
\caption{Ablation study on GAA module on generic subset of AVA dataset.  Best results are shown in \textbf{bold face}.}
\begin{tabular}{ccc|cc|cc|c}
\toprule
Visual characteristics & Spatial & Semantic & SRCC (mean) & PLCC (mean) & SRCC (Std. dev) & PLCC (Std. dev) & Accuracy \\
\midrule
\xmark                   & \xmark     & \xmark    & 0.701       & 0.704       & 0.265           & 0.272           & 80.00\%  \\
\cmark                   & \xmark     & \xmark    & 0.703       & 0.707       & 0.271           & 0.276           & 80.10\%  \\
\cmark                   & \cmark     & \xmark    & 0.704       & 0.707       & 0.265           & 0.272           & 80.19\%  \\
\cmark                   & \cmark     & \cmark    & \textbf{0.705}       & \textbf{0.710}       & \textbf{0.274}           & \textbf{0.281}           & \textbf{80.26\%}  \\
\bottomrule
\end{tabular}
\label{tab:abl_gaa}
\end{table*}

\begin{table*}[]
\centering
\caption{Subjects learned by the proposed model.}
\begin{tabular}{l|l}
\toprule
Category         & Learned subjects                                                                              \\
\midrule
Urban            & hair, head, graffiti, jacket, hat, wheel, street, line, eye, road                             \\
Seascapes        & people, sail, lighthouse, sailboat, surfboard                                                 \\
Portraiture      & eye, face, eyes, mouth, eyebrow, lips, glasses, teeth, neck, collar, beard, earring           \\
Sports           & helmet, hair, shoe, head, person, surfboard, wheel, tire, bike, sock, glove, wave, road, face \\
Birds            & head, beak, eye, face, ear, foot, eyes, mouth, nose                                           \\
All categories   & eye, hair, head, ear, nose, face, eyes, tail, mouth \\
\bottomrule
\end{tabular}
\label{tab:learned_subjects}
\end{table*}

\subsection{Ablation Study on the OAR Module}
\label{sec:ablation_study_oar}

Then we further investigate contributions of different information used by the OAR module for inferring object-level attentions. We consider 3 different settings for modeling object-level attentions, and compare these settings with the baseline model (denoted as Baseline in Table \ref{tab:abl_oar}) that directly predict image aesthetics with equally concatenated regional features. The first setting (denoted as Regional in Table \ref{tab:abl_oar}) predicts object-level attention-based on equally concatenated regional features, which is similar to the self-attention mechanism that learns the contribution of features according to the feature itself. In the second setting (denoted as Regional+Narrow global in Table \ref{tab:abl_oar}), not only equally-concatenated regional features are used, but also a narrow global feature is used, in the hope of global features can provide global information such as layout for determining regional contribution. The third setting (denoted as Regional+Wide global in Table \ref{tab:abl_oar}) further changes narrow global features to wide global features for predicting object-level attentions. The models have been evaluated on the generic subset of AVA dataset. 

We can observe three phenomena from the results. First, by introducing the object-level attention mechanism, even if global features are not introduced, the object-level attention mechanism can effectively increase the performance. We believe that this is because the object-level attention mechanism improves the representational power of the IAA model. Second, by further introducing narrow global features, the performance is further increased, which also supports our assumption that global feature can help to provide more information for determining contributions of different regional features. We believe that this phenomenon also aligns with the nature of human vision that it is hard for humans to judge which region is more attractive without looking at the overall structure and layout of the image. Third, by replacing narrow global features with wide ones, a performance gain can be observed, which implies spatially pooling MLSP features directly with global average pooling can lose necessary information and leaving more spatial information to the learning process can effectively increase the model performance.

\subsection{Ablation Study on the GAA Module}
\label{sec:ablation_study_gaa}

Experiments are also conducted to investigate contributions of different relationships mentioned in Sec. \ref{sec:gaa}, i.e. visual characteristic relationship, semantic relationship, and spatial relationship, in learning relevance between object pairs. Starting with the baseline with the OAR module but without the GAA module, we then introduce GAA modules with different relationship features for modeling graph attentions. The models have been evaluated on the generic subset of AVA. As it can be seen from the results (Table \ref{tab:abl_gaa}), the performance steadily improves as more information is introduced. This implies that the selected relationship is useful in determining pairwise relevance between OVCs.

\section{Further Discussion}

In this section, we attempt to interpret what has been learned by the proposed model from the perspective of the attention-subject consistency and the visual rightness theory. 
To this end, we firstly present some examples of resulting object-level attentions and graph attentions in Sec. \ref{sec:oaga}. Then we further investigate whether the proposed model has the ability to find subjects of different categories in Sec. \ref{sec:subject_area_pred}. 
In Sec. \ref{sec:objatt_vs_score} and Sec. \ref{sec:gatt_vs_score}, we study the relationship between object-level or graph attentions and aesthetic predictions. We then discuss how the results can be supported by attention-subject consistency and visual rightness theory.

\begin{figure}
  \centering
  \includegraphics[width=0.48\textwidth]{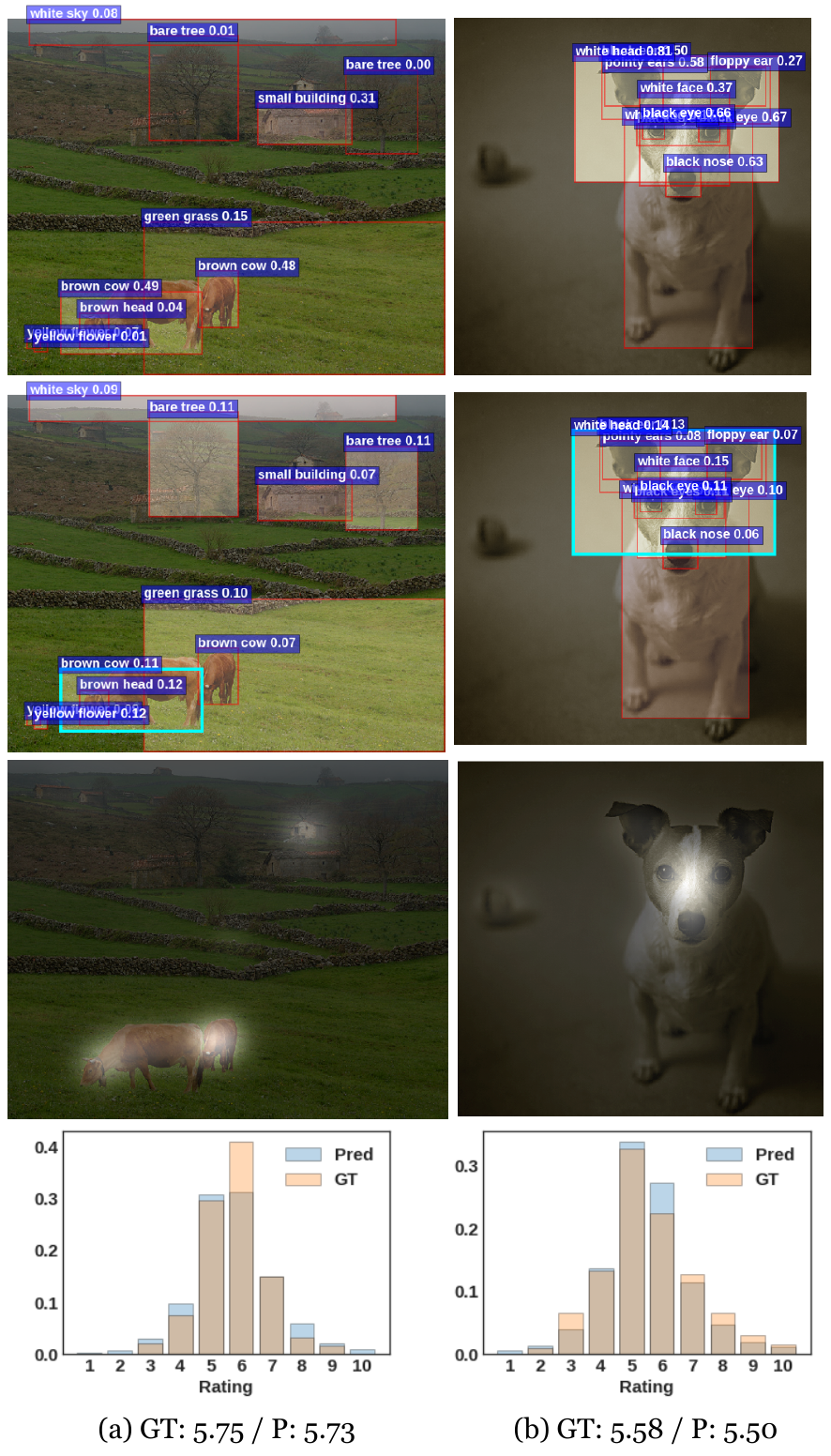}
  \caption{Examples for aesthetic predictions. Row 1: region-of-interests with projected IAA-driven object-level attentions; numbers in the labels are attention intensities, and higher brightness indicates a higher attention intensities. Row 2: graph attention maps for region with the highest object-level attention (labeled with cyan bounding boxes); Row 3: task-free pixel-level fixation maps which also use higher brightness for a higher attention level. Row 4: histograms for ground-truth (GT) and predicted (Pred) aesthetic rating distribution and the corresponding average scores.}
  \label{example_outputs}
\end{figure}

\subsection{Object-level Attention, Graph Attention, and Object Labels}
\label{sec:oaga}

One of the major flaws in most existing deep learning-based IAA methods is lacking interpretability. For the proposed framework, learned object-level attentions and graph attentions can be visualized or summarized for partially understanding what has been learned by the deep model from the data. We firstly show some examples of resulting object-level and graph attentions, and later sections (Sec. \ref{sec:subject_area_pred}, Sec. \ref{sec:objatt_vs_score} and Sec. \ref{sec:gatt_vs_score}) will conduct further explanation on the results. In Fig. \ref{example_outputs}, we present object-level attention maps (row 1 in Fig. \ref{example_outputs}) and graph attention maps (row 2 in Fig. \ref{example_outputs})). For object-level attention maps, the number in each bounding box label reflects the attention intensity of the corresponding region. The attention intensity is on a scale of $0\sim1$, and a higher value means stronger attention and higher contribution to the final ARD prediction. The attention intensities are projected to the corresponding bounded regions in ascending order so that low-attention regions will be covered by high-attention regions when there is any overlapping. For graph attention maps, we choose to visualize learned relevance between the region with the highest object-level attention (labeled with cyan bounding boxes) and other regions in Fig. \ref{example_outputs}, where the summation of graph attentions equals to 1 due to softmax. As shown, the object-level attention mechanism allows the contributions of different regions to be determined, comparing to the fixation maps only roughly highlight the most contributive regions. Additionally, the graph attention mechanism allow relevance between each pair of OVCs to be inferred for aesthetic predictions. In later sections, we summarize learned object-level attentions and graph attentions for model interpretation with the help of object class labels (e.g. `grass' in `green grass') and object attribute labels (e.g. `green' in `green grass') associated to each OVCs for model interpretations.

\subsection{Can Neural Network Learn What the Subject Is?} 
\label{sec:subject_area_pred}

According to the subject-attention consistency theory, we believe that the ability to determine subjects of images forms the basis of IAA. For different categories of photographs, we may expect some objects to be the subjects of that category. Thus, from training images whose predicted aesthetic scores are larger than 5 (regarded as high-aesthetic), we select objects whose intensities of object-level attentions on average are 0.04 greater than other objects as subjects. We do not select low-aesthetic images for finding common subjects based on attentions since the attentions on low-aesthetic images are usually deviated to non-subjects.

These selected attentive objects from high-aesthetic images are regarded as subjects recognized by the deep IAA model of the given category, and we show results from some of the categories (defined by semantic IDs given by AVA dataset) in Table \ref{tab:learned_subjects}. As the results shown in Table \ref{tab:learned_subjects}, the learned subjects can roughly characterize the content of the specific category. For example, in seascape, we would expect people with sailboats or surfboards as the main subjects of this category. 
We also find some objects receive strong attention across many categories. For example, we observe that objects related to facial features such as eyes, nose, face, and head (last row of Table \ref{tab:learned_subjects}) receive strong attention for many categories including portraitures, sports, birds, and some other categories unlisted in Table \ref{tab:learned_subjects}. We believe that this phenomenon arises from the fact that for photographs from different categories facial expressions and characteristics for both humans and animals are usually the main subjects to depict.

\begin{figure}
  \centering
  \includegraphics[width=0.48\textwidth]{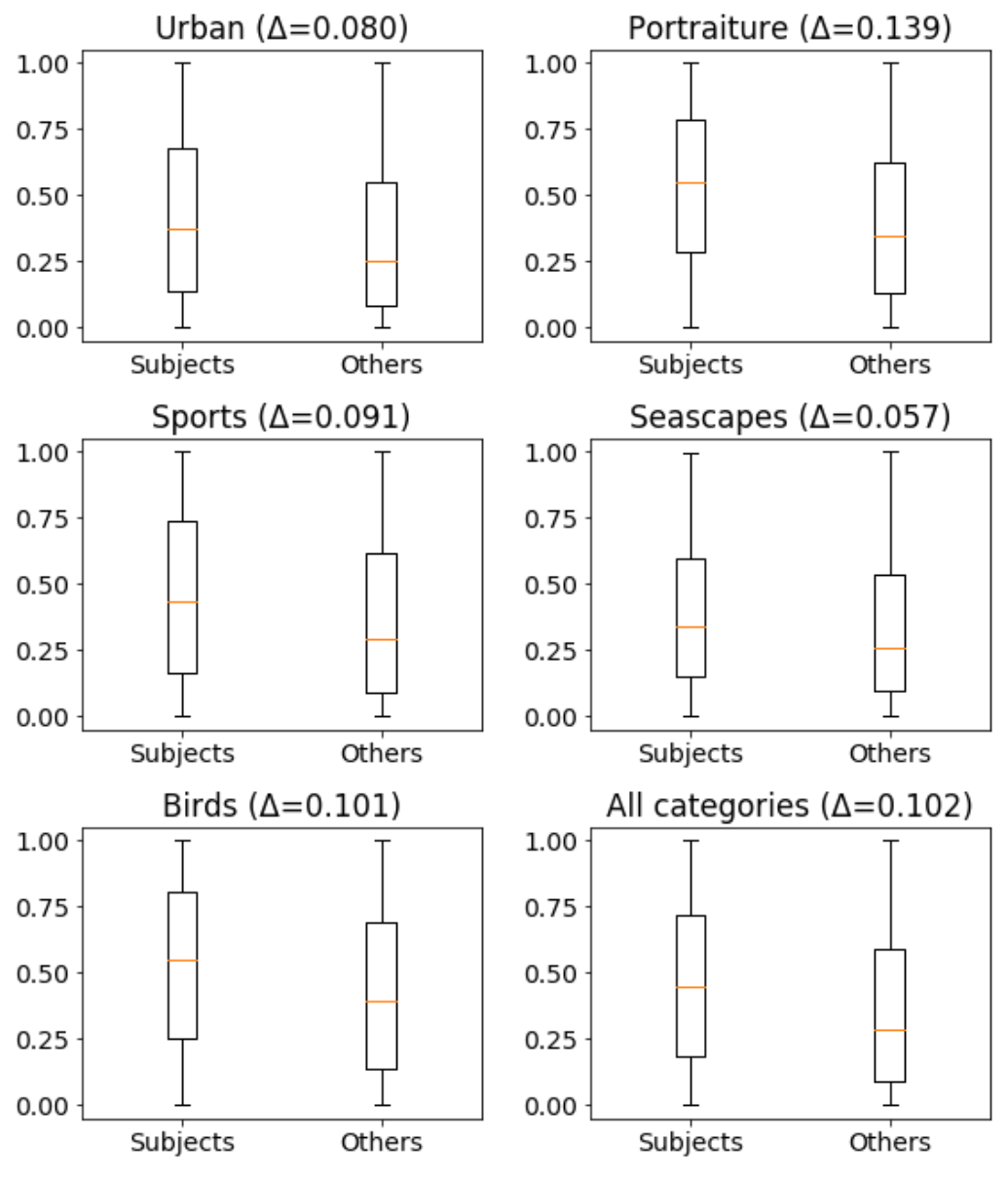}
  \caption{Boxplots for comparing distribution of object-level attention intensities between subject and non-subject regions. $\Delta$ is given by Eq. (\ref{eq:subject_att_diff}).}
   \label{fig:subjects_vs_others}
\end{figure}

Boxplots for comparing distribution of object-level attention intensities between subject and non-subject regions are given in Fig. \ref{fig:subjects_vs_others} with $\Delta$ is given by:
\begin{equation}
\label{eq:subject_att_diff}
    \Delta = \mu_s - \mu_o,
\end{equation}
where $\mu_s$ and $\mu_o$ are the mean of object-level attention intensities of `Subjects' set and `Others' set, respectively. A larger $\Delta$ suggests subjects can be more prominently seen from all OVCs by the deep IAA model, and therefore the deep IAA model has a higher capability to see the subjects for the category when $\Delta$ value is higher. 

\subsection{Object-level Attention vs. Aesthetic Scores} 
\label{sec:objatt_vs_score}

According to the subject-attention consistency theory, a photograph with a good composition can lead observers' attention to the subject of that image. In other words, if the composition cannot lead the attention of observers to the main subject, then the image has a bad composition.  And if that is the case, we should able to see two phenomena: 1) higher attention on subjects will have a positive effect on aesthetic predictions; 2) higher attention on non-subjects will have a negative effect on aesthetic predictions. Accordingly, if the attention of an object is positively correlated to aesthetic predictions, then the object is expected to be the subject; if the attention of an object is negatively correlated to aesthetic predictions, then the object is expected to be a non-subject.

\begin{figure}
  \centering
  \includegraphics[width=0.5\textwidth]{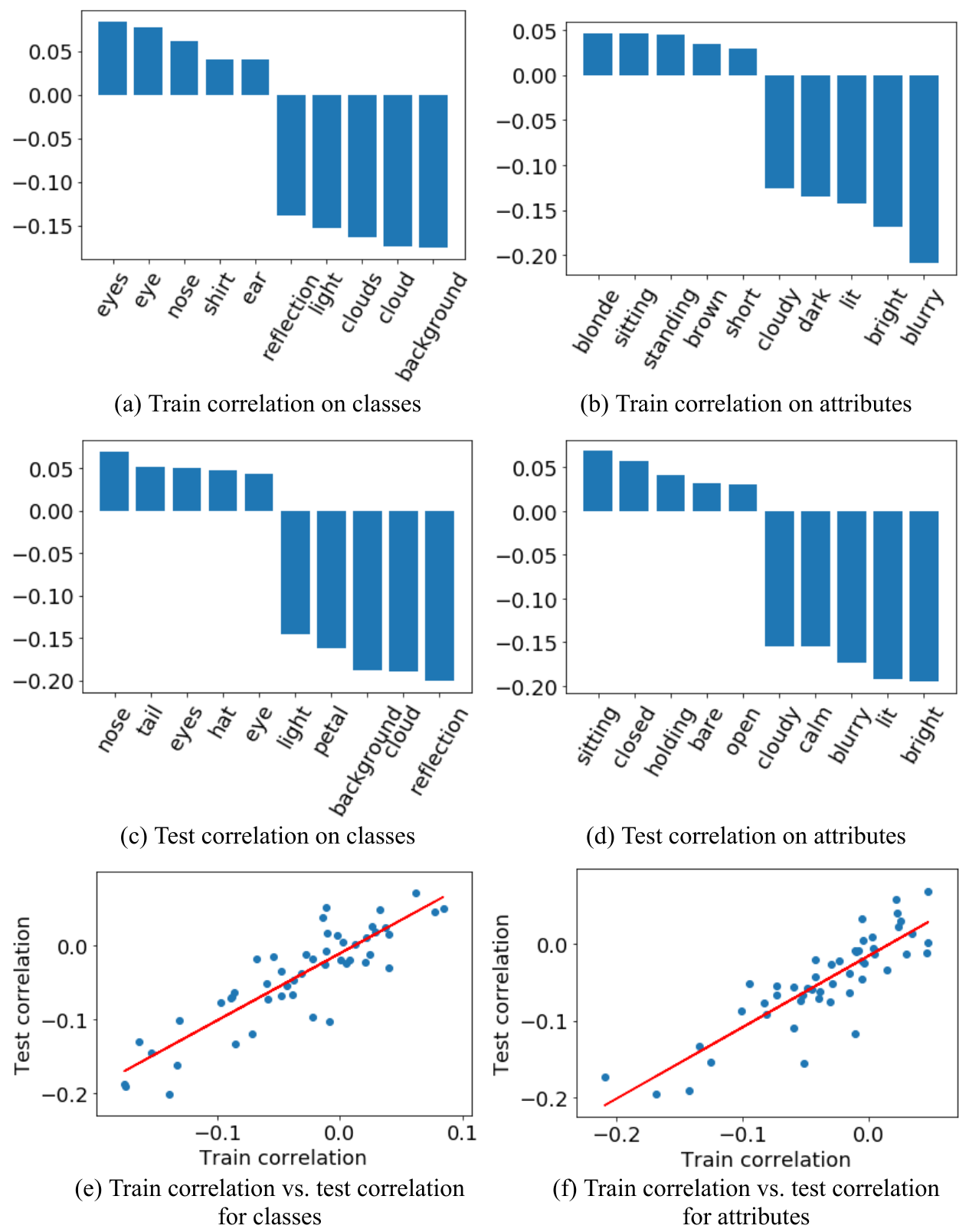}
  \caption{Correlations between object-level attentions and aesthetic scores.}
    \label{fig:unit_att_corr}
\end{figure}

We evaluate the correlation between object-level attentions for each object class (or attribute) and aesthetic scores of images that contain objects of the class (or attribute). For example, we select all images that contain `cloud' objects or (or `blurry' objects) and compute the correlation between the aesthetic scores of images containing `cloud' objects (or `blurry' objects) and intensities of object-level attentions on `cloud' objects (or `blurry' objects). Such computation has been conducted separately for the training set and the testing set, which results in different correlations for the training set and the testing set. We call these two correlations as train correlation and test correlation for simplicity. We select classes (or attributes) with top 50 frequencies in the training set for the investigation, and the results are shown in Fig. \ref{fig:unit_att_corr}. 

Fig. \ref{fig:unit_att_corr} shows object classes ((a),(c)) and attributes ((b),(d)) with top 5 and bottom 5 correlation on the training set and the testing set. As the results suggest, the strongest positive correlation only reaches around 0.05, which indicates there is not any class or attribute of objects whose attention could strongly improve the aesthetic scores. However, the absolute values of the negative correlations are much higher than the positive correlation. This implies that when attention has been driven to the bottom 5 objects, the aesthetic experience will be undermined. For example, for the train correlation on classes, `background' presents the strongest negative correlation, which means when attention has been direct to the background, the predicted aesthetic scores will be decreased. This supports our assumption that a bad composition could lead the attention of observers to deviate from the main subject, and therefore higher attention on non-subjects will result in lower aesthetic predictions. For the train correlation on attributes, `blurry' presents the strongest negative correlation, which implies when blurry regions receive higher attention, the aesthetic experience will be undermined. This also aligns with observations in previous works \cite{talebi2018nima, luo2011content} that image distortions such as blur are detrimental to aesthetic experience.

By comparing results of train correlation and test correlation, we could observe a slight difference between their orders of classes (or attributes) of correlation. To further check whether the learned rules will propagate from the training set to the testing set, we compute the correlation between the train correlation and the test correlation, and regression graphs are shown in  Fig. \ref{fig:unit_att_corr}(e) and Fig. \ref{fig:unit_att_corr}(f). We can see there is a strong correlation between the test correlation and the train correlation. This means that classes or attributes whose attention has a strong effect on aesthetic scores on the training set, will also strongly affect the results on the testing set.

\subsection{Graph Attention vs. Aesthetic Scores} 
\label{sec:gatt_vs_score}

According to the visual rightness \cite{locher2003empirical} theory in psychology, there must be a right organization of pictorial objects and such organization should be prominent to humans. In other words, a good composition should have a `salient' organization of objects, and the saliency on object organization can be partially reflected by learned graph attentions. Specifically, higher relevance implies the organization (modeled by visual characteristics, spatial and semantic relationships as discussed in Sec. \ref{sec:gaa}) of the object pairs is more salient for IAA. Thus, we further check whether any correlation between the intensities of graph attentions and predicted aesthetic scores can be learned by the deep IAA model. We conduct a correlation analysis similar to Sec. \ref{sec:objatt_vs_score} on the graph attentions of classes and attributes of OVC pairs. 

\begin{figure}
  \centering
  \includegraphics[width=0.5\textwidth]{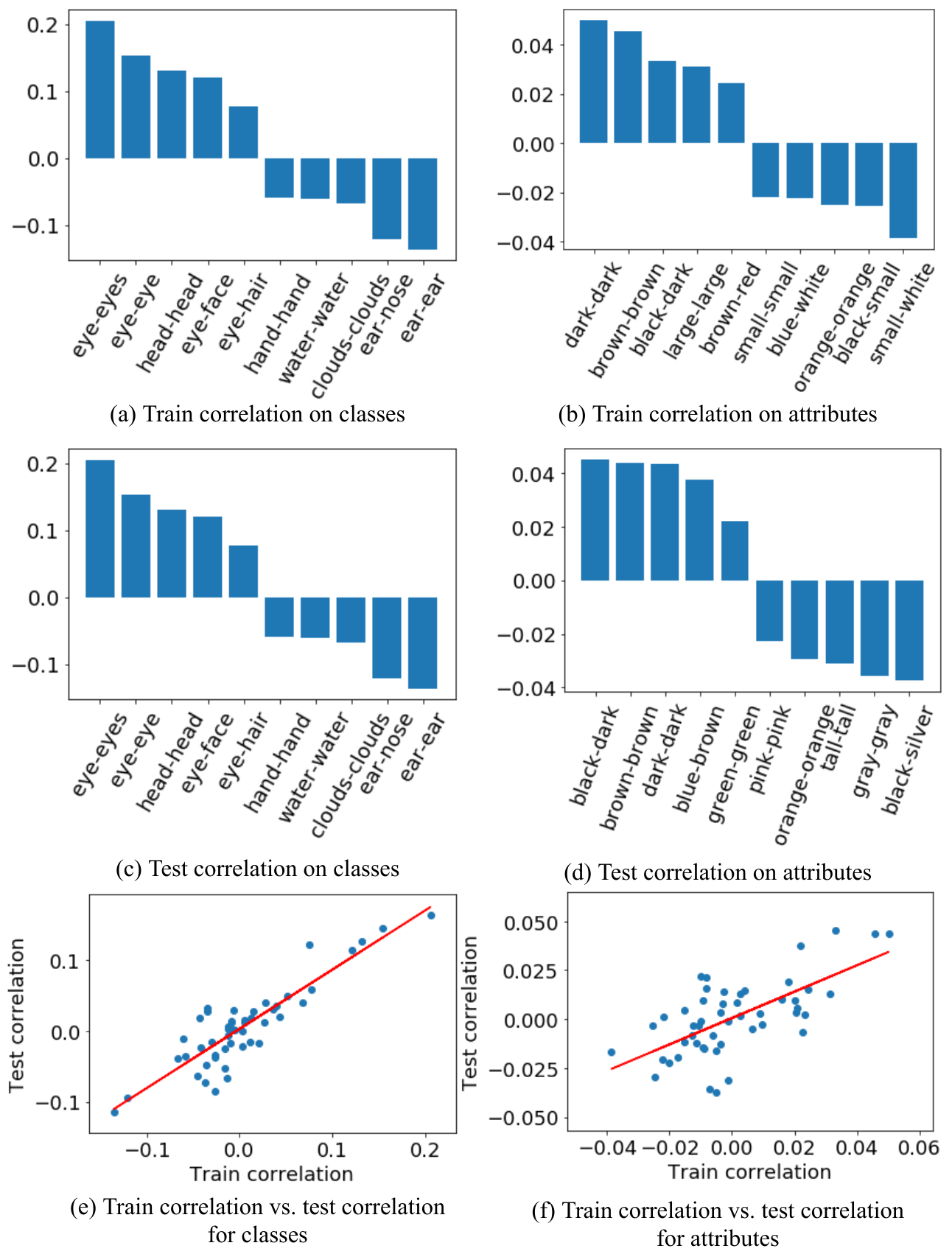}
  \caption{Correlations between graph attentions and aesthetic scores.}
    \label{fig:pair_att_corr}
\end{figure}

The results have been shown in Fig. \ref{fig:pair_att_corr}. As the results suggest, the correlation on object classes (Fig. \ref{fig:pair_att_corr}(a) and Fig. \ref{fig:pair_att_corr}(c)) are much stronger than the correlation on object attributes (Fig. \ref{fig:pair_att_corr}(b) and Fig. \ref{fig:pair_att_corr}(d)), indicating a prominent organization of object classes have a stronger effect on aesthetic scores than a prominent organization of object attributes. In Fig. \ref{fig:pair_att_corr}(a), class pairs with top 5 and bottom 5 correlation are mostly pairs of facial features. This implies the presence of prominent pairs of facial features will have a positive or negative effect on aesthetic scores. We believe that this observation is reasonable because facial expression and characteristics are usually the most compelling structure of an image. 
Results show that pairs that containing `ear' has a negative effect on aesthetic experience. We believe that this is also reasonable because when attention has been directed to peripheral parts of faces (i.e. pairs involve ears), it means that the central parts of the face (i.e. pairs involve eyes) of that image are less interested to the model, implying that the image has a bad composition.
Additionally, Fig. \ref{fig:pair_att_corr}(e) and Fig. \ref{fig:pair_att_corr}(f) also shows the effect of the saliency of object class (or attribute) pairs learned by the deep IAA model can be well propagated from the training set to the testing set.

\section{Conclusion}

In this work, we have proposed the Image Aesthetic Assessment from Object-level Visual Components (IAA-OVC) framework for image aesthetic assessment (IAA) inspired by photography and psychology theories. With the help of the proposed Object-level Attention-based Re-weighting module and Graph Attention-based Aggregation module, the proposed framework dynamically learns to aggregate features representing predefined object-level visual components and produces aesthetic representation accordingly. The framework allows regional contributions to be determined dynamically in a fine-granularity and characteristics to be aggregated across regions according to their relevance. Extensive experimental analysis over the most commonly-used AVA dataset demonstrates that our model is superior to existing relevant methods and ablation study confirms the effectiveness of the use of IAA-OVC. Finally, for the first time, this work has attempted to interpret how underlying object-level visual components are related to aesthetic predictions. Quantitative analysis for model interpretation suggest: 1) the deep model has gained the ability to find subjects of photographs of different categories via training; 2) evidences supported by the subject-attention consistency theory in photograph rules and visual rightness theory in psychology have been found from the results of the proposed model.


%





\ifCLASSOPTIONcaptionsoff
  \newpage
\fi



\bibliographystyle{IEEEtran}
\bibliography{IEEEexample}

\end{document}